\documentclass[]{raa}
\usepackage{graphicx,times}
\usepackage{natbib}
\usepackage{amssymb,amsmath}
\bibpunct{(}{)}{;}{a}{}{,}

\usepackage[a4paper=true,dvipdfm=true,pagebackref=true]{hyperref}
\hypersetup{pdftitle = The title of my PDF, pdfauthor = My name, pdfsubject= The subject, pdfkeywords = keyword1 keyword2 keyword3}
\hypersetup{colorlinks = true, linkcolor = green, anchorcolor = red, citecolor = blue, filecolor = red, pagecolor = red, urlcolor = red}

\begin{document}

   \title{Trends of stellar entropy along stellar evolution
}

 \volnopage{ {\bf 2012} Vol.\ {\bf X} No. {\bf XX}, 000--000}
   \setcounter{page}{1}

   \author{Marcio G. B. de Avellar\inst{1}, Rodrigo A. Souza\inst{1} and  J. E. Horvath\inst{1}
   }

   \institute{Instituto de Astronomia, Geof\'\i sica e Ci\^encias
Atmosf\'ericas - Universidade de S\~ao Paulo, Rua do Mat\~ao,
1226, 05508-900, Cidade Universit\'aria, S\~ao Paulo SP, Brazil;
{\it mgb.avellar@iag.usp.br} and {\it rodrigo.souza@usp.br} and {\it foton@iag.usp.br (JEH)}\\
\vs \no
   {\small Received 2012 June 12; accepted 2012 July 27}
}

\abstract{This paper is devoted to discuss the difference in the thermodynamic entropy budget {\it per baryon} in each type of stellar object found in Universe. We track and discuss the actual {\it decrease} of the stored baryonic thermodynamic entropy from the most primitive molecular cloud up to the final fate of matter in the black holes, passing through evolved states of matter as found in white dwarfs and neutron stars. We then discuss the case of actual stars of different masses throughout their {\it evolution}, clarifying the role of virial equilibrium condition for the decrease of the entropy and related issues. Finally, we discuss how gravity ultimately drives composition, hence structural changes along the stellar evolution all the way until the ultimate collapse to black holes, which may increase dramatically their entropy because of the gravitational 
contribution itself.
}

   \authorrunning{M.G.B. de Avellar, R.A. de Souza \& J.E. Horvath }            
   \titlerunning{Entropy in stellar evolution}  
   \maketitle

%
\section{Introduction}           
\label{intro}

Entropy, as defined by physicists, is a mathematical function that encodes the thermodynamic macro-state of a physical system constructed from the statistical description of that system, thus a thermodynamic potential from which all the physical quantities may be computed. The classical works of Gibbs and Boltzmann (see a review by \citealt{IngoMullerHistoryBook}) clarified the meaning of entropy and suggested that it is also related to the degree of disorder of a system, although only in a very restricted sense, then mainly related to the heat capacity analogies with solids, liquids and gases\footnote{Metals have lower heat capacities and lower entropies than liquids. At the same time, metals have less ways to spread out some injected energy through their internal structures which makes them more orderly than liquids.}. Entropy drives the thermodynamic evolution of a system in time; and it is related to the amount of energy, including heat, that is available to do work according to Clausius and other contemporaries (\citealt{IngoMullerHistoryBook}). See also a discussion about the minimization of energy and the maximization of entropy and the relation to the available free energy in \cite{IngoMuller2008}.

It is often stated that entropy plays a key role in any process in Universe, and by means of its study that we can achieve a better understanding of the fate of the Universe and its contents. The Universe contains several differentiated structures from the largest to the smallest scales, and it is precisely among the latter that we are going to study the relations of entropy in various evolutionary states of the most fundamental astrophysical objects in Nature: stars, beginning with molecular clouds all the way down to the (ultimate) formation of black holes.
The features of the thermodynamic entropy of a star along its main evolutionary phases will be addressed in Section \ref{hypotheses}. We present and discuss the results of entropy calculations in Section \ref{results}. We conclude in Section \ref{conclusions} by discussing how the whole stellar evolution is seen from the point of view of entropy considerations.


\section{Basic hypotheses and models}
\label{hypotheses}

\label{Timescales and equilibria in stellar evolution}
 Stars are essentially self-gravitating systems held by some internal pressure against collapse. The stability of stars is due to the fact
 that they spend almost all their lives in stationary states, in which the virial equilibrium relation $E_{pot} +2 E_{kin}=0$ among the various forms of potential $E_{pot}$ and kinetic $E_{kin}$ energy terms hold. During most of their ``active'' lives stars generate energy through nuclear reactions, either in its simplest form (hydrogen to helium conversion) or advanced versions (helium to carbon and beyond). A set of time-scales describing how the structural, thermal and energetic adjustments are made can be defined, including

\begin{itemize}

\item $\tau_{ff} \sim {\bigg({3\over{8 \pi G \bar{\rho}}}\bigg)}^{1/2}$ is the free-fall time-scale,
 an upper limit to the maximum velocity of propagation of any perturbation;

\item $\tau_{th} \sim {R^{2}\over{D_{th}}}$ where $D_{th}$ is the thermal diffusion coefficient
(ratio of the conductivity to thermal capacity), characterizing the time the star takes to
establish a stationary distribution of the temperature when the latter is perturbed;

\item $\tau_{KH} \sim {G M^{2}\over{R L}}$ is the Kelvin-Helmholtz time-scale, related to the time
it takes for the star to radiate of a significant fraction of the available energy;

\item $\tau_{nuc} \sim {\biggl( {1\over{X}}{dX\over{dt}} \biggr)}^{-1}$
related to the burning of a given nuclear fuel with mass abundance $X$.
\end{itemize}

Stars in steady state satisfy $\tau_{ff} \, < \, \tau_{th} \, < \, \tau_{KH} \, < \, \tau_{nuc}$.
Whenever a nuclear fuel is exhausted, the last inequality is violated, and the star seeks a
new equilibrium state by contracting on a Kelvin-Helmholtz time-scale. Note that, because of
the high thermal content and the first inequality, this is {\it not} actually a collapse. However, in
each of the stages gravity gets stronger, and ultimately drives changes in the state of
matter inside (i.e. degeneracy). Therefore, we can state that entropy should be produced
but also radiated away (to the envelope and later away from the star), while hydrostatic
equilibrium is maintained. In the formation of
a star, or late in the final true collapse (supernova stage), this entropy generation and
radiation is even more marked, since the contraction is much less ``gentle'' and
out-of-equilibrium, therefore irreversible processes play a major role.

We are thus lead to consider the Second Law of Thermodynamics in a familiar form for the
system star + environment, namely the equation \ref{eqn1}

\begin{equation}
\frac{dS}{dt} = \Sigma - \oint \overrightarrow{J_{S}} ~ d\overrightarrow{\Pi}
\label{eqn1}
\end{equation}

In principle, tracking all the sources of entropy inside the star ($\Sigma$) and the flux of
the entropy currents $\overrightarrow{J_{S}}$ throughout the boundary $\Pi$ we could calculate
the increase or decrease of entropy for each stage {\it of evolution of a given star of a given mass}. Instead of that, one can just calculate
initial and final states, thus ``weighting'' the relative importance of both terms on the
right hand side of the above equation. Note that there could be ejection of mass (and entropy with it) in some explosive stages, although we shall not discuss the details of this complication in the remaining of the discussion, since we will compare just the entropy content in the final configurations of the objects under study (these final configurations will have the same number of baryons, for the reasons explained later). Besides, as we shall see later in Figure \ref{thermoEntropy}, a considerable amount of entropy is lost with the processes that lead to the explosive stages. In addition, the luminosity equation for a differential shell to evolve whenever there is a compression or
expansion reads (in Lagrangian coordinates)

\begin{equation}
\frac{dL}{dm} = \epsilon - T\frac{dS}{dt}
\label{eqn2}
\end{equation}

with $\epsilon$ the nuclear energy generation rate and $S$ the entropy per mass unit and we have neglected the energy loss in the form of neutrinos. As we see, there could be luminosity generation even without nuclear reactions, provided there is enough temporal variation of the
enclosed entropy. Processes like ionization driven by compression, for example, contribute to the second term in some stages in between steady burning stages. The second term of Equation \ref{eqn2}, also known as the thermal term, is more important for giant branch stars, not discussed here.

A couple of comments are necessary regarding the above equation since it bears a very important meaning when one has to compute the structure of a star. If one turns off the term for nuclear energy generation, the virial theorem guarantees that the star will contract somewhat releasing part of its gravitational energy to compensate the decrease of its internal energy. However, even if one does not ``turn off'' the source term, nuclear reactions slowly change the composition and the temperature gradients inside the star and then the structure, leading it to a new gravitational configuration as the star expands or contracts (depending on the internal energy balance).

These gravitational adjustments imply that a gravitational work is done on the stellar matter which, in turn (due to the energy sources) drives an  exchange of heat between adjacent shells of stellar matter. In this way, the above equation is a direct consequence of the principle of conservation of energy: $dQ/dt = dU/dt + P dV/dt$. Thus, the change of entropy with time is a consequence of the very process of evolution of the star, i.e., the attempt of gravity to sustain the star in a state of (quasi) hydrostatic equilibrium, a heat-exchange process among matter shells (that is why some authors call this term, not very precisely, the ``gravitational energy source''). Therefore, this term is related to $\tau_{KH}$, and since the stars evolve on a much slower time-scale than $\tau_{KH}$, except for the star on the Hertzprung gap, $TdS/dt \simeq 0$ and the condition of (local) thermal equilibrium is satisfied. Any complete and realistic model including transient adjustments must take into account the $TdS/dt$ term. However, when computing the structure of a star we realize that we have to make some assumptions about the initial and boundary conditions and, in the case of a collapsing or expanding phase, the models may depend on these conditions, sometimes quite strongly. Here we meet the real physical meaning of the equation just described: one cannot compute the structure of a star without knowing its previous history as stated in \cite{ClaytonBook}.

As a matter of fact, it should be remembered that the above eq. \ref{eqn2} does {\it not} determine the luminosity of a star as seen by a distant observer. That equation is prescription of what the luminosity should be in order to maintain the energy balance, e.g., the nuclear fusion and gravitational adjustments balance the energy losses, and is valid in any differential shell inside the star . The energy outflow from the star is ultimately determined by the radiation transport mechanisms such as diffusion, convection and conduction, all dependent on the shape and value of the temperature gradient, to be determined as a solution of the full set of structure equations.

After all these considerations we shall, in order to be able of compare the different evolutionary phases of our model stars, first set a conserved quantity. The baryon number is precisely tailored for such a purpose, and we shall fix it to the value $N=1.61\times10^{57}$ throughout this work, unless explicitly stated. This is not an arbitrary choice: the number
corresponds to a mass (for small binding energy) of about $\sim \, 1.35 \, M_{\odot}$. It corresponds rather well to a solar-type example while hydrogen burning is considered, and even beyond, and is very close to the critical transition mass between the ``evolutionary phases'' from the high-mass tail of the mass distribution of white dwarfs to the low-to-average mass tail of the mass distribution of neutron stars as well.

Our program to discuss the whole changes of the entropy will be the following: taking into account that a star will end up as one of the three kinds of compact objects (a white dwarf, a neutron star or a black hole),\footnote{Because the fate of a star depends on the mass it has at the moment it enters in the main sequence, the progenitors of the compact objects must have different initial masses. Roughly, a white dwarf has a progenitor with $1-7.5~M_{\odot}$, a neutron star has progenitor with $8-25~M_{\odot}$ and a black hole has progenitor with $\geq 25~M_{\odot}$.} we set these final configurations as having the the same number of baryons (our conserved quantity), $N=1.61\times 10^{57}$, and we calculate the thermodynamic entropy of each compact object. After that we track back what would be the progenitor of each of our compact objects, e.g., what is the mass of the main sequence star that produced the compact objects with $1.61\times 10^{57}$ baryons. We calculate the entropy of these stars in some chosen epoch of their lives during the Main Sequence. And then, we track back the entropy of the primordial clouds that produced these stars in the main sequence.

Our assumptions on the stellar state at each stage are:

\begin{itemize}
\item the total energy is given by $E_{tot}=E_{int}+E_{kin}+E_{pot}$;
\item the virial condition $E_{pot}=-2\times E_{kin}$ is satisfied;
\item the components have equilibrium particle distributions, for example, equipartition of energy holds for ideal gases, $E_{kin}\sim kT$.
This is a justified as long as the thermal timescale remains very short, as is usually the case.
\end{itemize}

In general, the entropy is a function of the (internal) energy, the volume and the number of particles of a system: $S=S(E_{in},V,N)$. Thus, we must properly choose the physical models that yield the energy and volume in each stage for a fixed number of baryons. The evolutionary stages we will discuss are:

\begin{enumerate}

\item {\bf White dwarf:} White dwarfs are the endpoint of the evolution of ordinary stars with $\sim1$ to $\sim7~M_{\odot}$.

From the point of view of stellar evolution, things depart considerably from the previous stages of the lives of ordinary stars. At this stage hydrogen can no longer burn in the star and the core contracts under its own gravity after achieving the Sch\"onberg-Chandrasekhar condition (\citealt{SchonbergChandrasekhar1942}) changing the hydrostatic equilibrium. This is the end of the Main Sequence stage (see below) and the star moves out of it entering into a completely new path in the H-R diagram. For a low- to intermediate mass star like the one we are dealing with, the contraction of the core proceeds until the point it is eventually halted by the degeneracy pressure of the electron gas, while the conservation of energy plus the virial relation together force the envelope to expand. After a series of structural changes, including the ignition of helium in degenerate conditions in the core (helium flash), the star will eject the outer envelope in a series of thermal pulses of increasing amplitude. Mass loss at this stage is very large and can not be ignored, therefore to hold the baryon number fixed as before we are not considering the actual evolutionary path but rather an ideal model situation for the sake of clarity. This is the end point of the evolution of this star because the remnant cannot generate energy. It will cool down releasing all the thermal energy it had stored.

It is widely known that the actual composition of a ``typical'' white dwarf is mainly carbon-oxygen (C-O) and they have typical masses of about $0.6~M_{\odot}$. However, in order to conserve the baryon number of $1.6\times10^{57}$ we had to assume the creation of a very heavy white dwarf with $\simeq 1.35M_{\odot}$ well to the tail of the mass distributions of these stars. The composition is then different: it is probably oxygen-magnesium-neon (O-Mg-Ne) white dwarf. To produce such a heavy white dwarf, the progenitor star was chosen to have mass of $7~M_{\odot}$, roughly $8.3\times10^{57}$ baryons, which lose $6.7\times10^{57}$ baryons during the mass loss phase, which is a reasonable assumption.

Thus, in order to calculate the structure of this star we assumed the ultra-relativistic regime with a polytrope of index $n \sim 3$, since the mass of this object is quite near the Chandrasekhar limiting mass (\citealt{Chandrasekhar1931}). The equation of state is then given by $P = K\rho^{4/3}$. The assumed central density is $1\times 10^{10}g/cm^{3}$ from which $R_{WD}\simeq 0.002 R_{\odot}$ ($\sim 1500 km$) is obtained.

We calculated the entropy of the WD in two specific moments: the hot initial phase, after the thermal pulses, where the core temperature is $T_{HWD}\sim 5\times 10^{8}K$ and for a very late and evolved phase, when the core temperature is $T_{CWD}\sim 1\times 10^{5}K$.

\item {\bf Neutron star:} Neutron stars are are formed by the collapse of a massive star $8-25 M_{\odot}$ resulting in a compact object of $\geq 1.2 M_{\odot}$ and $R \sim 10 km$\footnote{Another possible way to form a neutron star is via the accretion-induced collapse of a white dwarf (\citealt{vandenHeuvel2011}).}. Neutron stars are supposed to have all the same composition since the burning stages of the evolution reach the limit of the iron, from where no more exothermic processes are possible. So, with iron as the starting point for further evolution, the core contracts to a completely new phase of (dense) matter.

Neutrons stars also have a mass distribution, possibly two peaked (\citealt{valentimRangelHorvath2011}), and the different masses are possibly due to the masses of the progenitor stars and the mass loss processes during earlier stages of evolution during the post main sequence phase. Thus, in order to produce a neutron star with $1.6\times10^{57}$ baryons, near to the lighter part of the mass distribution, we assumed a progenitor with $11~M_{\odot}$ that lose something about $11.5\times10^{57}$ baryons in the ultimate supernova explosion.

We calculated the entropy of our neutron star in three distinct moments: the hot proto-neutron star phase whose temperature was assumed to be $T\sim 5\times 10^{11}K$ and $R_{PNS}\sim 55.75 km$ (this is about five times the radius of the forthcoming neutron star, due entirely to the ``hot phase''); a later ``stationary'' hot phase with temperature of $T = 1\times 10^{9}K$ and $R_{NS}\simeq 11.15 km$, which settles a few hours after the formation at most; and a final cold ``stationary'' phase with temperature of $T = 1\times 10^{7}K$ and $R_{NS}\simeq 11.15 km$ representing the cooling of the isolated neutron star after $\sim 10^{6} yr$ approximately.

\item {\bf Black hole:} We end our calculations with the ultimate state of the collapsed matter, the black hole. The actual formation of this extreme compact object is marked by the death of a very massive star ($\geq 25M_{\odot}$). In a similar way as with the neutron star formation, we assumed a progenitor with $25M_{\odot}$ that, after its normal evolution, ejects $28.1\times10^{57}$ baryons ending with the formation of a $1.6\times10^{57}$ baryons black hole. After the formation of the event horizon, the final object emits thermal radiation at a Hawking temperature $T_{BH} = \frac{\hbar c^{3}}{2k_{b}\pi G(Nm_{u})}\simeq 1.8\times 10^{-7} K$. The celebrated proportionality between the entropy and the area (\citealt{Bekenstein1973,Bekenstein1974}) now applies, since all forms of matter have disappeared beyond the horizon.

\item {\bf Main sequence star:} We follow the entropy evolution, dominated by the ideal gas component, using Townsend's MadStar online tool (http://www.astro.wisc.edu/$\sim$townsend/) to create ordinary main sequence star models, burning hydrogen to helium and solar metallicity. We created four main sequence stars: one with $1.35~M_{\odot}$ that will evolve in some 4 billions of years into an old ordinary star (it will evolve further to a C-O white dwarf that we will NOT study here); one with $7~M_{\odot}$ that will evolve in some hundreds of millions of years to a O-Mg-Ne white dwarf; one with $11~M_{\odot}$ that will evolve in some dozens of millions of years to a  neutron star; and one with $25~M_{\odot}$ that will evolve in some hundreds of thousands of years to a black hole. The MadStar is a online tool based upon an approach by Bill Paxton on the famous Eggleton code. Although some limitations exist, none of them have strong influence to our purposes.

\item {\bf Molecular cloud:} Modern determinations of stellar forming conditions (\citealt{CaproniAbrahamVilasBoas2000}) have shown the occurrence of
substantial clumping within molecular clouds. We consider the formation of stars inside these clumps which have typical temperatures $\sim 20 K$, typical masses $\sim 0.2M_{\odot}$ and typical densities $\sim 10^{5}cm^{-3}$ that merge together to form a single star. The radius of each small cloud is then $R_{SMC}\simeq 8.23\times 10^{16} cm \simeq 0.03 pc$. Thus, each small cloud amounts to $1.19\times 10^{56}$ baryons. The clumps at the immediate stage of star formation are opaque to radiation, but their entropy is largely dominated by the ideal gas component (we neglect magnetic fields in this discussion). The merging of a few of these small clumps will produce a star with the assumed baryon content, in a complex process driving the star towards the Zero-Age Main Sequence (ZAMS) immediately following the ignition of hydrogen and the establishment of the hydrostatic equilibrium condition.

Recalling, we want to study, at the same time, the thermodynamic entropy of a given $1.6\times10^{57}$ baryons in different states of matter, i.e., in different degrees of compactification and the changes of the entropy content along the the evolution of the objects that originated these final compacted baryons in the first place. As mentioned above, we have three compact stars representing three different exotic states of matter. We need then four original molecular clouds: three for the three compact stars and another one to account for $1.6\times10^{57}$ baryons enclosed in a ordinary star in ``normal'' state.

Clumping of small molecular clouds add up to form four molecular clouds, with $1.6\times10^{57}$ baryons, $8.3\times10^{57}$ baryons, $13.1\times10^{57}$ baryons and $29.7\times10^{57}$ baryons, respectively, that will form the four main sequence stars that, in turn, will form our final four objects whose entropy will be studied.

\end{enumerate}

It is important to check for each stage of evolution the state of degeneracy, since degenerate gases follow a different entropy expression than ideal gases. In Table \ref{tab1} we give the Fermi temperature in each stage, remembering that the degeneracy occurs if $T_{object}<< T_{F}$.

\begin{table}[h!]
\caption{Fermi temperatures [K] of each evolutionary stage given by our hypotheses: molecular cloud (MC), ordinary Main Sequence (MS) star, white dwarf (WD), proto-neutron star (PNS), and neutron star (NS). It is important to notice that in the table below we calculated an average Fermi temperature employing the average density of each object. The Fermi temperature is given by $T_{F}=\frac{1}{k_{b}}\frac{\hbar^{2}}{2m_{u}}\Big(3\pi^{2}\eta\Big)^{2/3}$, where $\eta=\frac{N}{V}$, generally a function of the radial coordinate $r$.}
\centering
\begin{tabular}{|c|c|c|c|c|c|c|c|c|}
\hline
 & MC & Star0 & Star1 & Star3 & Star4 & WD & PNS & NS \\
\hline
$T_{Fe^{-}}$ & $\sim 10^{-7}$ & $\sim 10^{5}$ & $\sim 10^{5}$ & $\sim 10^{5}$ & $\sim 5\times 10^{4}$ & $\sim 10^{11}$ & $\sim 10^{14}$ & $\sim 10^{15}$ \\
$T_{F_{H^{+}}}$ & $\sim 5\times 10^{-11}$ & $\sim 150$ & $\sim 120$ & $\sim 70$ & $\sim 30$ & $\sim 5\times 10^{7}$ & $\sim 10^{10}$ & $\sim 10^{12}$ \\
\hline
\end{tabular}
\label{tab1}
\end{table}

Employing the values in Table \ref{tab1} we can calculate the entropies in all the stages.


%
%
%
%
%


\section{Results and Discussion}
\label{results}

In each subsection below we discuss how we calculate the components of the entropy and its total value in each evolutionary stage. Then, in Table \ref{tab2} we show which term of the entropy is dominant. This is of course impossible for a black hole, which does not have any ordinary component left and needs a separate consideration.

A comparison with other known entropy sources in the Universe is interesting, as discussed by \cite{Frampton2009}. It is important to remark that stars giving rise to neutron stars/black holes represent $\sim 1 \%$ of the $10^{22}$ stars present in the visible universe.
In any case, the entropy content of all the stellar populations is tiny compared to other known components (i.e. CMB photons) and thus irrelevant
for the whole budget.

\subsection{Molecular cloud}

Giant molecular clouds are the main cradle of stars. Fragmentation of a giant cloud and further clumping and collapse of smaller units form the main blocks, as discussed above. These are composed mainly by neutral molecular hydrogen, being described roughly by the ideal gas law. Here we assume that the small clouds described in $\S$ \ref{hypotheses} clump together to amount the determined number of baryons as described in the previous section and that each is in equilibrium just before the clumping and collapse to form the stars in the zero-age main sequence (ZAMS). Because the temperature is $T_{MC}= 20K \gg T_{Fe^{-}/ions}$ there is no degeneracy. Then, the entropy can be calculated by the expression for an ideal gas:

\begin{equation}
\label{cloudEntropy}
S_{MC}=\sum S_{SMCbaryons}=\sum N_{SMC,b}k_{b}\Bigg(ln\Big(\frac{V_{SMC}}{N_{SMC,b}}\Big)+\frac{3}{2}ln\Big(\frac{E_{inSMC}}{N_{SMC,b}}\Big)+const\Bigg),
\end{equation} where $const=\frac{3}{2}ln\Big(\frac{4\pi m_{u}}{3h^{2}}\Big)+\frac{5}{2}$, $k_{b}$ is the Boltzmann constant, $V_{SMC}$ is the volume of each small cloud and $m_{u}$ is the atomic mass unit. The important assumption here is that the clouds are composed by molecular hydrogen only. 

The total entropy in each of our four cases is the sum of the entropy of a certain number of small clouds that add up to a specific number of baryons. So, for our first object, 6.73 small clouds add up $1.6\times 10^{57}$; for the second, 34.98 small clouds add up $8.3\times 10^{57}$; for the third, 54.97 small clouds add up $13.1\times 10^{57}$; and for the fourth, 124.93 small clouds add up $29.7\times 10^{57}$. Recall that each small cloud has $1.19\times 10^{56}$ baryons, as described in the previous section.

From our assumptions: $E_{tot}=0$, $E_{pot}=-2E_{kin}$ and $E_{kin}=\frac{3}{2}Nk_{b}T_{cl}$, we finally find

$$
S_{MC1}=4.77\times 10^{42} erg/K \hspace{0.25cm}or\hspace{0.25cm} S\rightarrow \frac{S}{k_{b}N} = 21.47;
$$

$$
S_{MC2}=24.8\times 10^{42} erg/K \hspace{0.25cm}or\hspace{0.25cm} S\rightarrow \frac{S}{k_{b}N} = 21.65;
$$

$$
S_{MC3}=39.0\times 10^{42} erg/K \hspace{0.25cm}or\hspace{0.25cm} S\rightarrow \frac{S}{k_{b}N} = 21.57;
$$ and

$$
S_{MC4}=88.6\times 10^{42} erg/K \hspace{0.25cm}or\hspace{0.25cm} S\rightarrow \frac{S}{k_{b}N} = 21.62.
$$

The calculation of the entropy of a molecular cloud in an earlier stage is tricky and requires a careful consideration (not attempted here).
In the transparent stages of the cloud, the radiation is not effectively coupled to matter and it is not clear whether it should be included. Nevertheless, this stage happens well before any actual condensation stage and is not important for our considerations.

\subsection{Main Sequence stars}

The second evolutionary stage encompass the Main Sequence (MS) stars, the region in the Hertzprung-Russell (HR) diagram where the stars stay most of their lives. For stars in the range of masses worked out here, $1.35~M_{\odot}$ to $25~M_{\odot}$, the period of residence in the MS is $\sim 1/M^{2.5--3}$, while the energy generation is mainly due to the so-called $p-p$ chain for star with masses up to $2~M_{\odot}$ and due to the CNO cycle for masses above $2~M_{\odot}$. In our case, this is roughly $4 \, Gyr$. As stated, typical temperatures of our four models are $\geq 10^{7}K$ while the Fermi temperature is $\sim 10^{5}K$ for electrons and $\sim 10^{2}K$ for ions of hydrogen ($H^{+}$). Therefore, there is essentially no degeneracy along the Main Sequence, except maybe for a small degree of degeneracy in the inner core. 

The four molecular clouds collapsed to form four models of Main Sequence stars with ZAMS masses $1.35~M_{\odot}$, $7~M_{\odot}$, $11~M_{\odot}$, and $25~M_{\odot}$. The MadStar evolutionary code calculates the entropy of the structure in a way that the entropy of each star at Main Sequence is:

$$
S_{1.35M_{\odot}}=8.26\times 10^{44} erg/K \hspace{0.25cm}or\hspace{0.25cm} S\rightarrow \frac{S}{k_{b}N} = 3718;
$$

$$
S_{7M_{\odot}}=53.5\times 10^{44} erg/K \hspace{0.25cm}or\hspace{0.25cm} S\rightarrow \frac{S}{k_{b}N} = 4671;
$$

$$
S_{11M_{\odot}}=78.3\times 10^{44} erg/K \hspace{0.25cm}or\hspace{0.25cm} S\rightarrow \frac{S}{k_{b}N} = 4331;
$$

$$
S_{25M_{\odot}}=212\times 10^{44} erg/K \hspace{0.25cm}or\hspace{0.25cm} S\rightarrow \frac{S}{k_{b}N} = 5173.
$$

The general behaviour of entropy with ageing inside the MS is to get smaller and smaller, as illustrated with the example of the star with $1.35~M_{\odot}$:

$$
S_{ZAMS} = 8.26\times10^{44} erg/K \hspace{0.25cm}or\hspace{0.25cm} S\rightarrow \frac{S}{k_{b}N} = 3718;
$$

$$
S_{t = 0.8~Gyr} = 8.13\times10^{44} erg/K \hspace{0.25cm}or\hspace{0.25cm} S\rightarrow \frac{S}{k_{b}N} = 3659;
$$

$$
S_{t = 2~Gyr} = 7.70\times10^{44} erg/K \hspace{0.25cm}or\hspace{0.25cm} S\rightarrow \frac{S}{k_{b}N} = 3466;
$$ and

$$
S_{t = 4~Gyr} = 5.10\times10^{44} erg/K \hspace{0.25cm}or\hspace{0.25cm} S\rightarrow \frac{S}{k_{b}N} = 2295.
$$

A more general consideration of the behaviour of the entropy inside the Main Sequence for several masses and in the Post-Main Sequence stages will be given in section \ref{actualStars}.

\subsection{White dwarf}

The condition of matter inside a white dwarf is fairly different in the two stages we considered. In the hot phase, the electrons are degenerate, but the ions constitute basically a Boltzmann gas. Therefore we have three components for the total entropy. While we can still use the same terms in the expression \ref{cloudEntropy} for the ions and use the expression $S_{rad}=\frac{4}{45}\frac{\pi^{2}k_{b}^{4}}{c^{3}\hbar^{3}}VT^{3}$ for the radiation, we need a new expression for the entropy of the degenerate matter:

\begin{equation}
S_{WDe^{-}}=\frac{1}{2}\frac{\pi^{2}(x_{e}^{2}+1)^{1/2}Nk_{b}\Big(\frac{k_{b}T_{WD}}{m_{e}c^{2}}\Big)}{x_{e}^{2}},
\end{equation} where $x_{e}\equiv\frac{p_{fe^{-}}}{m_{e}c}$ and $p_{fe^{-}}$ is the Fermi momentum of the electron sea. The factor $1/2$ comes from the supposition that $N_{p}=N_{n}=N_{e}$, that is reasonable for a white dwarf. From our calculation $S_{e^{-}}\sim 10^{40} erg/K$ and $S_{ions}\sim10^{42}erg/K$.

Thus, for the hot initial state we obtain

$$
S_{WDhot}\simeq 1.27\times 10^{42} erg/K \hspace{0.25cm}or\hspace{0.25cm} S\rightarrow \frac{S}{k_{b}N} = 5.72.
$$

However, as the star cools down the electrons remain degenerate, but the ions suffer a phase transition to form a {\it Coulomb lattice} (\citealt{MestelRuderman1967}). The Debye temperature marking the crossover of these regimes is for our model $\theta_{D}\simeq 1.8\times 10^{8}K$. Because we chose a final ``stationary'' state with a temperature $\sim 10^{5}K<<\theta_{D}$, we are well inside the regime where the heat capacity goes with $\sim T^{3}$ corresponding to phonon lattice excitations. The entropy is given then by

\begin{equation}
\label{WDcoldions}
S_{WDcold-ions}=\frac{16N\pi^{4}k_{b}}{15}\Bigg(\frac{T}{\theta_{D}}\Bigg)^{3}\sim 10^{33} erg/K
\end{equation} while for the degenerate electrons $S_{e^{-}}\sim 10^{36}erg/K$.

For the cold final state we then obtain

$$
S_{WDcold}\simeq 4.14\times 10^{36} erg/K \hspace{0.25cm}or\hspace{0.25cm} S\rightarrow \frac{S}{k_{b}N} = 1.86\times10^{-5}.
$$

We see that the entropy budget changes along the cooling age of a white dwarf. At first, in the hot WD state, most of the entropy is stored in the ions. In the cold WD state, the degenerate electrons eventually hold the largest fraction of entropy, being the excess of entropy carried away by the photons.

\subsection{Neutron star}

Again, the state of matter inside the star differs radically in these phases. In the proto-neutron star phase the assumed temperature is $5\times 10^{11}K$ resulting in electrons which are still degenerate (the Fermi temperature in this configuration is $T_{Fe^{-}}\sim 10^{13}K$), but the neutrons can still be considered a non-degenerate gas with some degeneracy correction (the Fermi temperature in this state is $T_{Fions}\sim 5\times 10^{10}K$).

Then, the entropy of a proto-neutron\footnote{Here we used a radius $R_{PNS}= 5 R_{NS}$, where $R_{NS}=11.15 km$; which mimics the hot
stage before cooling} star is given by the entropy of degenerate electrons plus the entropy of a neutron Boltzmann gas:

$$
S_{PNS}\simeq 2.96\times 10^{42} erg/K \hspace{0.25cm}or\hspace{0.25cm} S\rightarrow \frac{S}{k_{b}N} = 13.32.
$$

After the proto-neutron star phase, the neutron star settles and quickly cools down via neutrino emission. Our model for cold neutron stars results from solving the Tolman-Oppenheimer-Volkoff equations (\citealt{FridolinBook}) complemented by the SLy4 equation of state (\citealt{DouchinHaensel2001}), a particularly suitable choice for a compact star composed by very neutron rich matter with interactions. We selected the star with $N = 1.6 \times 10^{57}$ baryons which correspond to a (gravitational) mass $M \simeq 1.23 M_{\odot}$ and radius $R=11.15 km$, showing the effects of stronger gravity in these objects through a lower total mass (larger binding).

Assuming a temperature of $T_{NS}\sim 1\times 10^{9}K$ in this later stage, the neutrons become degenerate and support the star against further collapse. Then the stellar entropy is given by the entropy of degenerate neutrons (we have neglected the small fraction of electrons/protons enforced by beta equilibrium):

\begin{equation}
S_{NS}=\frac{\pi^{2}(x_{n}^{2}+1)^{1/2}Nk_{b}\Big(\frac{k_{b}T_{NS}}{m_{n}c^{2}}\Big)}{x_{n}^{2}},
\end{equation} where $x_{n}\equiv\frac{p_{fn}}{m_{n}c}$ and $p_{fn}$ is the Fermi momentum of the neutron sea.

The entropy is then

$$
S_{NShot}\simeq 1.21\times 10^{39} erg/K \hspace{0.25cm}or\hspace{0.25cm} S\rightarrow \frac{S}{k_{b}N} = 5.45\times10^{-3}.
$$

As the neutron star cools down, eventually reaching a temperature of $T_{NS}\sim 1\times 10^{7}K$ in the core, its entropy decreases further to

$$
S_{NScold}\simeq 1.21\times 10^{37} erg/K \hspace{0.25cm}or\hspace{0.25cm} S\rightarrow \frac{S}{k_{b}N} = 5.45\times10^{-5}.
$$

At this stage, the entropy decrease is mainly due to photon emission.

\subsection{Black hole}

A black hole is a region in space-time in which an event horizon has been formed and enclosed all the matter of the progenitor, and ultimately it does not matter which kind of particles contributed to it.  Because there is no access to the interior content of a black hole, a thermodynamic description of the collapse cannot be based on the entropy of the contents since these are lost from the observable universe.

The black hole entropy depends solely on the observable properties of the black hole: mass, electric charge and angular momentum. Because of the area theorem (\citealt{Bekenstein1973,Bekenstein1974}), these three parameters appear in a combination defining the area. The expression for the entropy of a Schwarzschild (non-rotating, uncharged) black hole is given by:

\begin{equation}
S_{BH}=\frac{k_{b}A}{4G\hbar}=\frac{4k_{b}\pi G(Nm_{u})^{2}}{\hbar c},
\end{equation} where we used the fact that $A=4\pi R_{H}^2=16\pi (GM/c^{2})^{2}$ with $R_{H}=2GM/c^{2}$.

The numeric value of the entropy of $1.6\times10^{57}$ baryons enclosed by the event horizon is then

$$
S_{BH}\simeq 2.63\times 10^{61} erg/K \hspace{0.25cm}or\hspace{0.25cm} S\rightarrow \frac{S}{k_{b}N} = 1.2\times10^{20}.
$$

That is many orders of magnitude larger than the preceding states of matter. This can be considered a measure of the degree of irreversibility of the ultimate collapse to form black holes out of known matter and energy, that is, the entropy currents are quickly stopped from flowing out the black hole because of the horizon formation and the highly irreversible collapse generates huge amounts of entropy. This qualitative picture, however, cannot be taken too seriously, since it is not clear yet how exactly the entropy is located in the area (making $S$ a non-extensive quantity), a subject of much discussion and calculations for string theorists and loop quantum gravity researchers (\citealt{GhoshPerez2011,JMR2005}) beyond the scope of our work.

\subsection{Thermodynamic entropy summary}

\begin{figure}[!h]
 \centering
 \includegraphics[scale=0.82, angle=0]{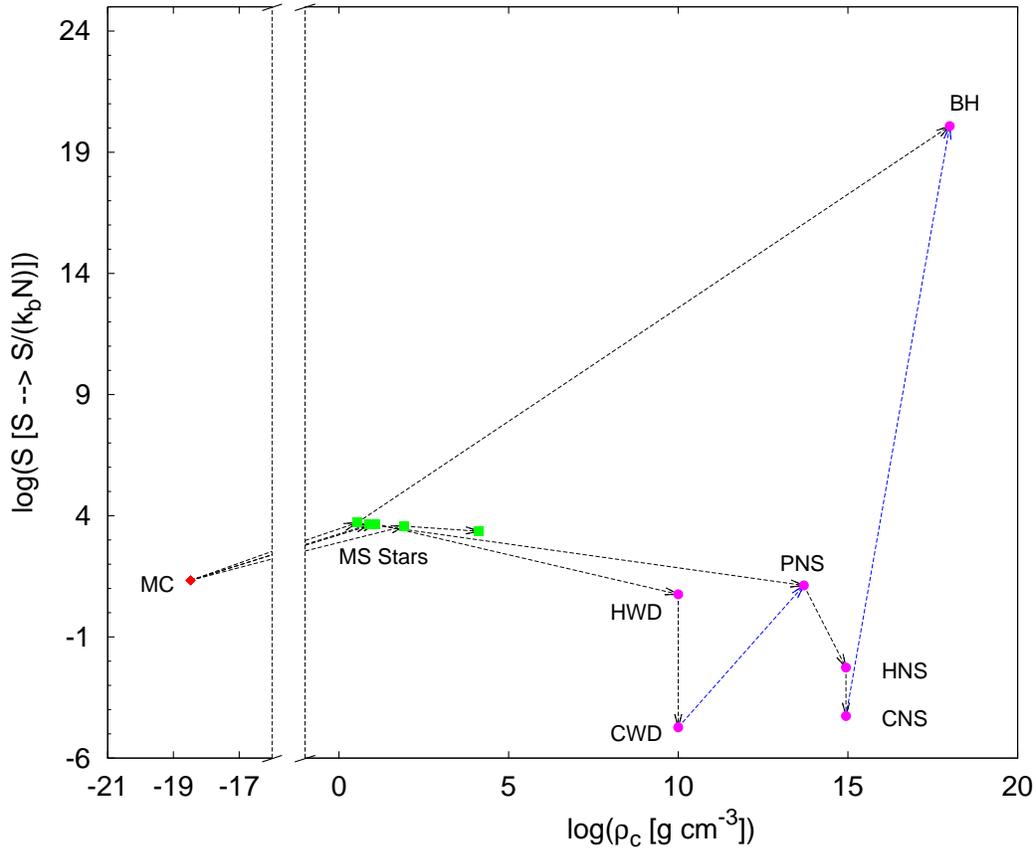}
\caption{Thermodynamic entropy {\it per baryon} in units of $k_{b}$ {\it versus} the central density of the objects in each stage of evolution. The (red) diamonds are the four molecular clouds; the (green) squares are the four main sequence stars (note that there are five squares, but one of them represents an evolved version of the same $1.35~M_{\odot}$ star); and the (pink) circles represents the compact stars (white dwarf, neutron star and black holes plus the proto-neutron star). The (black) arrows show the changes (or ``evolution'') of the entropy as stellar evolution proceeds. The two (blue) arrows going from the cold white dwarf to the proto-neutron star and from the cold neutron star to the black hole are the special case of induced collapse. Notice that for a black hole we assumed an effective central density of $10^{18}g/cm^{3}$ for plotting purposes}
\label{thermoEntropy}
\end{figure}

In Table \ref{tab2} we show the dominant entropy contribution in each compact object with $1.6\times10^{57}$ baryons reflecting, {\it at the same time}, how entropy end up stored in the final configurations at different states of matter and the evolutionary path as the stars go through their lives. In Figure \ref{thermoEntropy} we show the thermodynamic entropy {\it per baryon} in units of $k_{b}$ as a function of the central density at each stage of the evolution (notice that for a black hole we assumed an effective central density of $10^{18}g/cm^{3}$ for plotting purposes).

It is most important to realize that in stellar evolution gravity drives ultimately the changes in the energetic processes in the interior of stars and the final states of matter in the compact objects accordingly to the mass of each progenitor star.

The important comparison is then between the progenitor main sequence stars of $7~M_{\odot}$, $11~M_{\odot}$ and $25~M_{\odot}$, and the ``initial'' final configurations hot white dwarf, hot neutron star (passing through the proto-neutron star phase), and black hole. We clearly see a trend in lowering the entropy from earlier stages until the final configurations.

From the point of view of the baryon content in relation to the state of matter inside the compact objects, we see that although the main sequence star of $7~M_{\odot}$ has lower entropy than the $11~M_{\odot}$ star, the entropy jump to a state of lower entropy is higher from the star of $11~M_{\odot}$ to the neutron star than from the star of $7~M_{\odot}$ to the white dwarf. Thus, for the same number of baryons, the more compact object has the lower entropy and this difference is due to the structural changes of matter, from ordinary matter in white dwarfs to very neutron-rich matter in neutron stars. In this respect the entropy of the proto-neutron star is very similar to the entropy of the hot white dwarf, since in broad terms, the proto-neutron stars is a kind of iron white dwarf, with matter in a state that resembles the state of white dwarf.

Gravity is the ultimate force driving the entropy changes and stellar evolution. Then one can wonder why the most compact object, the black hole, has a entropy that is so many orders of magnitude higher than the other compact objects with the same number of baryons, apparently contradicting the very conclusion we stated in the previous paragraph. The answer lays in the gravity field itself and its putative entropy content. We shall see below that the latter could nicely explain the big difference of entropy content between low and high curvature stars.

\begin{table}[h!]
\caption{Entropy components [erg/K] of each compact object with $1.6\times10^{57}$ baryons given by our hypotheses: white dwarf (WD), proto-neutron star (PNS), neutron star (NS), and black hole (BH).}
\centering
\begin{tabular}{|c|c|c|c|c|c|c|c|}
\hline
 & Radiation & Ideal baryons & Ideal electrons & Degenerate $e^{-}$ & Crystal & Degenerate $n$ & Area \\
\hline
HWD & $\sim 10^{37}$ & $\sim 10^{42}$ & -- & $\sim 10^{40}$ & -- & -- & -- \\
CWD & $\sim 10^{26}$ & -- & -- & $\sim 10^{36}$ & $\sim 10^{33}$ & -- & -- \\
PNS & $\sim 10^{42}$ & $\sim 10^{42}$ & -- & $\sim 10^{42}$ & -- & -- & -- \\
HNS & $\sim 10^{32}$ & -- & -- & -- & -- & $\sim 10^{39}$ & -- \\
CNS & $\sim 10^{26}$ & -- & -- & -- & -- & $\sim 10^{37}$ & -- \\
BH & $\sim 10^{-17}$ & -- & -- & -- & -- & -- & $\sim 10^{61}$  \\
\hline
\end{tabular}
\label{tab2}
\end{table}

\begin{figure}[!h]
 \centering
 \includegraphics[scale=0.80, angle=0]{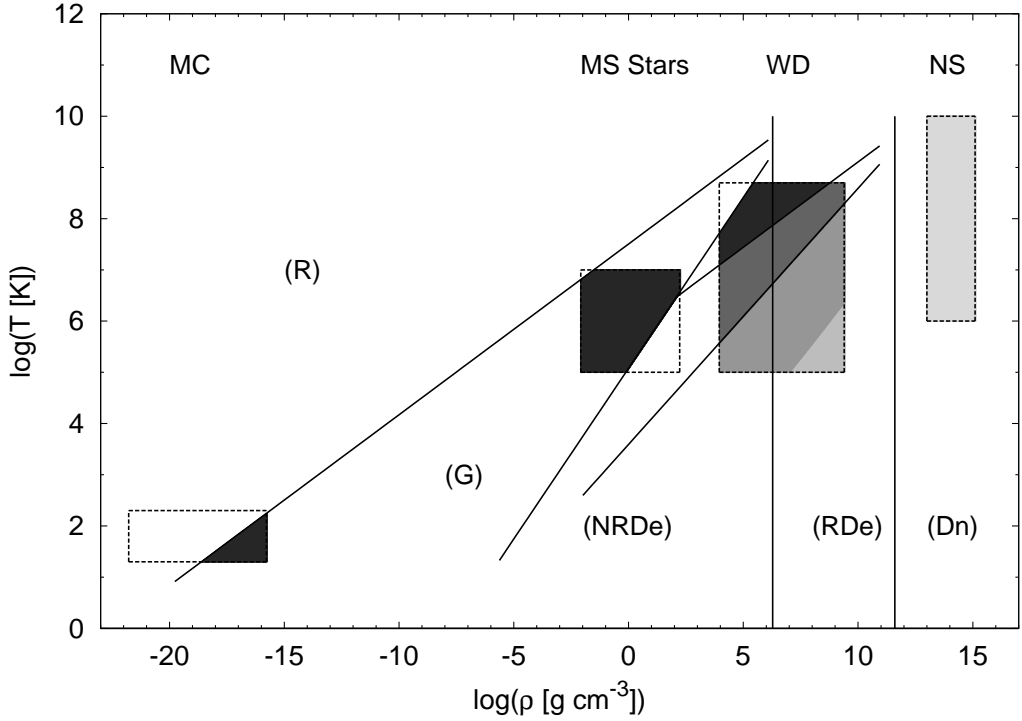}
\caption{The region above the first left line (R) is radiation pressure-dominated; the region between the first left line and the second line (G) is gas pressure-dominated; the pressure of non-relativistic degenerate electrons (NRDe$^{-}$) is dominant in the region between the second line and the first vertical line; the region between the two vertical lines is dominated by the pressure of the relativistic degenerate electrons (RDe$^{-}$); the region to the right of the second vertical line is dominated by the pressure of degenerate neutrons (Dn). The other two lines, the ones crossing the first vertical line, define a change in the regime of the specific heats of the ions inside white dwarfs due to crystallization. Regarding entropy, in the darkest regions the entropy is dominated by the entropy of the Boltzmann gas; in the second darker region the entropy is dominated by the entropy of a hot lattice of ions; in the third darker region the entropy is dominated by the entropy of cold crystallized ions lattice; in the lightest grey region of the WD box the entropy is dominated by the entropy of degenerate electrons; and finally, in the region of the NS box the entropy is dominated by the entropy of degenerate neutrons.}
\label{planoTxRHO}
\end{figure}

To finish the thermodynamic entropy summary, we show in Figure \ref{planoTxRHO} regions in the plane {\it T vs $\rho$} which entropy regime is dominant in the typical density and temperature range corresponding to the models studied here. We also show the well-known dominant pressure regimes for comparison.

We see that the entropy regimes are {\it not} directly connected to the pressure regimes, although they overlap in many situations. This can be seen in the specific case of a white dwarf case: while the electron degeneracy pressure is dominant through the whole range of densities of white dwarfs, the entropy of the degenerate electrons is dominant only in the regime of low temperatures and high densities (the lighter grey region of the WD box of Figure \ref{planoTxRHO}). This change is directly related to the evolution of the white dwarfs, more specifically to the cooling and to the fact that there is no energy generation inside the core of these objects. A similar behaviour may occur for the degenerate neutron composition case, and it is worth further study.

\subsection{Actual Main Sequence stars and beyond}
\label{actualStars}

The considerations made above about the entropy of an idealized path for a fixed baryon number were intended to show how this quantity evolves as successive stellar equilibrium states arise. However, it is clear that actual stars do {\it not} complete the whole evolution, and therefore it is important to discuss how the entropy behaves for the whole range of stellar masses in actual stars.

It is currently agreed that nuclear reactions start for the central conditions of stars with mass above $0.08 M_{\odot}$ (for solar composition). Besides, the stellar structure is stable up to a high mass limit of at least $90 M_{\odot}$ and possibly of $\sim 130 M_{\odot}$. Within these two extremes, the stellar structure varies and the entropy characterizing the star is subject to a quite different behaviour. Generally speaking, it is well-known that a stellar core or shell may transport energy in the {\it radiative} mode (diffusion of photons driven by a small temperature gradient, the so-called local thermodynamic equilibrium condition, LTE) or, when the stability criterion formulated in terms of the size of the temperature gradient $\frac{dT}{dr}$ is violated, in the {\it convective} mode, in which large-scale motions of the fluid itself are involved (\citealt{KippenhahnBook}). The latter regime is difficult to model since a variety of scales and complex phenomena (i.e. turbulence) play a crucial role. Therefore, simple approaches ignoring the multi-scale dimension in favour of a representative ``bubble'' are employed, the most popular being the mixing length theory which parametrizes a number of physical effects with a single dimensionless quantity $\alpha$, expected to be $O(1)$. Even the plain substitution of the true temperature gradient by the adiabatic temperature gradient ${dT\over{dr}}\big|_{ad}$ is many times employed for calculations based on the fact that the actual value of the gradient cannot be large on physical grounds, since convection is very efficient and there is little room for a large value build-up of temperature differences (see the textbooks in \cite{CarrollOstlieBook,KippenhahnBook}).

The behaviour of entropy in both cases (diffusive vs. convective) is very different, and an additional condition, that of virial equilibrium, is determinant for its value, at least as long as the balance is dominated by the ideal gas component and not by degenerate electrons or radiation, as we shall see in a moment.

The virial relation states that, in equilibrium, the gravitational and thermal energies satisfy $\frac{3}{5}\frac{GM^{2}}{R}=2\frac{3}{2}NkT=3\frac{M}{\mu m_{H}}kT$, where $N$ is the total number of particles, $\mu$ is the mean molecular weight of the gas and $m_{H}$ is the hydrogen mass. If we assume a constant value of the density, a simple manipulation of the relation yields $T \sim \frac{G\mu m_{H}}{k} M^{2/3} \rho^{1/3}$. In such case, the expression of the ideal gas shows that the entropy decreases with increasing temperature, in complete agreement with the Second Law {\it when subject} to the virial equilibrium  condition. It is said sometimes that the star ``digs an entropy hole'' along its evolutionary path, but this behaviour ceases as long as some other component (i.e. degenerate electrons) dominates the entropy balance.

The observation that large-scale motion of the fluid is the dominating transport mechanism beyond a certain value of the temperature gradient (the convective regime) leads to a variety of situations along the stellar evolution for a given mass, and also to different configurations for stars with different masses. But another important feature of the stellar interior will be relevant for an overview of the entropy. It is related to the condition in which nuclear energy is released by fusion reactions. Simple calculations show that around $\sim 2 M_{\odot}$ the main reactions from {\it CNO} catalytic cycle overcome the so-called {\it p-p} channel, and therefore the heavier stars burn hydrogen in a much more dramatic way, since the latter is {\it very} dependent on the temperature (in contrast with the former which is very mildly dependent on $T$).

Thus, stars below this ``great divide'' threshold develop a steep gradient only at the outer layers, whereas above it the steep gradient is present in the core. Therefore, the structure of stars in the first regime (termed {\it Lower Main Sequence}) is radiative on the inside, and convective outside, all the way to the photosphere. Those in the {\it Upper Main Sequence}, in turn, become convective inside and keep a radiative envelope.

If we consider stars of lower and higher masses away from the threshold, there are two important boundaries to be noticed in the mass parameter: stars below $\sim 0.3 M_{\odot}$ are totally convective as the result of the systematic inwards advance of the convective envelope from $\sim 2 M_{\odot}$ solar mass towards the lowest values. In addition, the convective core {\it grows} with mass at the other end, i.e., $M_{convective}/M_{total}$ grows from $\sim50\%$ for a $20~M_{\odot}$ star up to $\sim80\%$ for a $100~M_{\odot}$ star. Those stars in between present both convective and radiative regions, as discussed above.

\begin{figure}[!h]
 \centering
 \includegraphics[scale=0.60, angle=0]{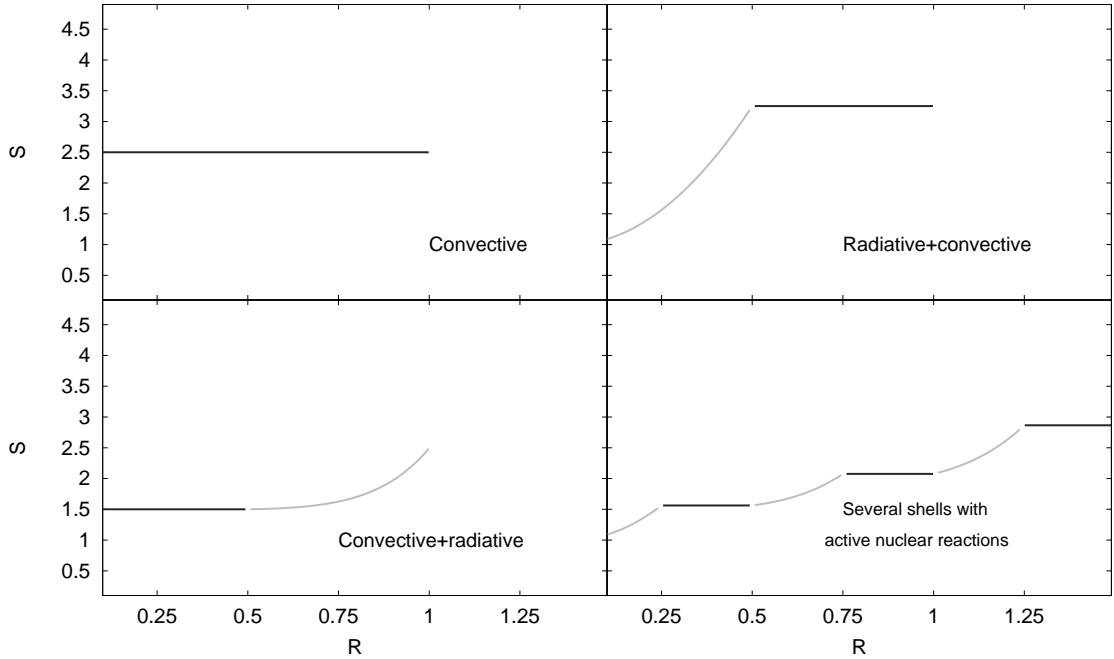}
\caption{Entropy regimes (arbitrary units). For plotting purposes we used the same scale for all plots.}
\label{entropyRegimes}
\end{figure}

Given these features and considering the homogenization provided by convection, the entropy profile of any star residing in the Main Sequence can be depicted as in Figure \ref{entropyRegimes}. The first extreme (below $\sim 0.3 M_{\odot}$) corresponds to a constant value of the entropy, as showed in the upper left panel of Figure \ref{entropyRegimes}. More massive stars, of the solar type, and up to around $\sim 2 M_{\odot}$ are represented by the upper right panel of the same Figure, featuring a radiative core and a convective envelope. Still more massive ones, up to $\sim 20 M_{\odot}$  or so, correspond to the profile shown in the lower left panel; and those stars above this value are again represented by a constant value of $S$ (upper left panel).

From the point of view of the entropy, similar behaviour occurs when stars leave the Main Sequence evolving towards the right of the HR diagram. Low-mass stars develop an increasingly massive helium core, mostly degenerate for solar-mass stars and below. The envelope expands and becomes convective inwards (the reverse of the Hayashi track in the stellar formation process...), and the entropy distribution resembles the one in Figure \ref{entropyRegimes} (lower right panel). Note that degenerate cores, by their own nature, do {\it not} diminish their entropies with increasing temperature. In contrast, upper main sequence stars develop normal, non-degenerate cores and depending on their exact value, end their lives as white dwarfs (those up to around $8 M_{\odot}$) or ignite further nuclear reactions. The full sequence of available nuclear cycles (including true fusion reactions and photo-disintegration rearrangements, yielding net energy) is achieved for masses  beyond $\sim 10 M_{\odot}$. For the range $8-10 M_{\odot}$ it is expected that a degenerate $O-Mg-Ne$ core is formed, but the conditions would not be enough to go beyond the carbon cycle. Since when each of the combustion cycles are no longer possible, the inert core contracts and the reactions formerly at the center migrate to a shell around it, the so-called ``onion structure'' develops for the more massive stars, in particular, in those with $M \geq 10 M_{\odot}$ which would complete the ignition of all available nuclear reactions. The entropy, in turn, adopts a constant value inside convective cores, but given that the virial condition must be maintained, its value {\it decreases} for each new cycle. Meanwhile, the radiative shells adopt the entropy distribution growing from the inner to the outer edge of the shell in each case. Thus, the entropy distribution for any given stage of massive stars is qualitatively similar to that shown in the lower right panel of Fig. \ref{entropyRegimes}. This process of decreasing the entropy per baryon at the center is more pronounced for the lighter range of massive stars, since the relation $T \sim \frac{G\mu m_{H}}{k} M^{2/3} \rho^{1/3}$ leads to $\frac{T^{3}}{\rho} \propto M^{2}$, and the ideal gas entropy then increases with increasing mass, being much higher for the range $15-20 M_{\odot}$ and beyond. Since the end of these stars as core collapse supernovae depend on the binding energy of the core, directly related to the entropy per baryon, the low entropy per baryon in the $\sim 10 M_{\odot}$ range was considered as a favourable condition for the explosions to succeed, although many other factors seem to be at play in this process (\citealt{MurphyDolenceBurrows2013}).

At last, a remark should be made regarding actual stars and systems of stars. Binary evolution channels are very important for the formation and evolution of compact objects, but introduce new complications to the entropy considerations which lie beyond the scope of the present approach. For comprehensive considerations of a possible common envelope phase we refer to \cite{ge01,ge02}.

\section{Conclusions}
\label{conclusions}

We have seen that the gravitational compactification of stellar matter makes the thermodynamic entropy of matter to {\it decrease} from the main sequence stars to their correspondent compact stars in an ordered, monotonic sequence. This decrease of the thermodynamic entropy of matter is not at all in contradiction with the Second Law since the gravitational contraction releases high-entropy radiation (and even neutrinos) in a way that more than compensates the decrease in the matter entropy of the object. Thus, it is the ability to radiate away entropy what keeps the entropy of stars decreasing.

Starting with the condensations in a molecular cloud, as the collapses goes on, the core of the object is heated by the contraction up to the point the fusion reactions of protons into helium begins. At this moment, the processes in the core balance gravity and the contraction halts, but the thermodynamic entropy of matter remains lower than that it was in its initial state. Later on, as the nuclear fuel of the star is exhausted, a new contraction begins, being halted only when the heat capacity becomes positive again when a phase transition of matter occurs. After a few billion years, a white dwarf is formed with a still lower thermodynamic entropy.

In other words, the entropy decrease of matter with the gravitational contraction is at the expense of an increase in the entropy of the immediate environment due to the release of very high-entropy radiation/neutrinos during the life of stars in the Main Sequence and beyond. There is a delicate balance of two processes. The fusion reactions {\it per se} can be thought to lower the entropy because the number of particles diminishes. However, this process is highly exothermic, and increases the local entropy due to local heating. Which effect dominates depends on the local temperature: at low temperature (relative to nuclear scales), the entropy gain from the exothermic reactions is favoured. When the temperature of the core reaches some threshold, the reactions become entropically unfavoured and stop. This coincides with the jump to the next stage of the evolution of the objects. In the white dwarf and neutron star stages, the entropy budget suffers a stronger influence from cooling processes, since the structure of these objects will remain unchanged for time-scales which are infinite in practice. However, the injection of entropy in the environment is never too extreme, for example, a (type II) massive star supernova will produce around $3 \times 10^{42} erg/K$ of entropy mainly in neutrinos. A hundred times this figure will be injected due to the dissipation of the massive ejecta, ultimately leading to dust heating of the interstellar medium. These numbers are still lower than the total entropy introduced by the heating of dust due to the absorption of ordinary radiation of stars (\citealt{Bousso2007,Frampton2009}).

Gravity plays a major role in all these contractions and entropy changes, driving stellar evolution towards its end. Gravity is responsible, again, for the formation of a black hole. As seen above, collapse under gravity causes entropy to increase enormously; however here the black hole itself ``cools down'' to a very low Hawking temperature (for a solar-mass or so black hole), much smaller than the external temperature while its entropy increases astonishingly. Because black holes have negative heat capacity, they absorb radiation faster than they can emit by the Hawking mechanism. We face a very special case: the environment is {\it cooled} by the collapse (differently than in normal stars) while the entropy of the matter in the black hole is very high (\citealt{Wallace2010}).

As mentioned earlier, the gravitational field itself may carry and store entropy. This entropy has been related to the curvature inside the stars, but it is negligible from Newtonian objects like the molecular clouds, main sequence stars and white dwarfs. For neutron star, constructed under the framework of General Relativity, the entropy of gravity is expected to be larger, but still does not alter significantly the entropy budget of this very compact star. However, things change considerably when a massive main sequence star collapses to form a black hole.
Black holes are ``pure gravity'', a singular point in the metric that in which the curvature term goes to infinity, forming an object from which no matter can ever escape. That is why this collapse actually {\it increases} the entropy stored in this object.
There is considerable activity in the community to construct and characterize a gravitational entropy. A definite recent example is the 
proposal of \cite{CET2013} in which a ``gravitational entropy'' has been suggested based on the Bel-Robinson tensor which
makes use of the Weyl part of the curvature tensor $C_{abcd}$ (see \cite{SL2014} for applications). Thus, such a proposal attaches an increasing entropy to the gravitation which is related to the increasing curvature value. As a particular example, the gravitational entropy reduces to the Bekenstein-Hawking value for the case of a Schwartzchild black hole. Therefore, if this proposal stands, the origin of the black hole huge entropy could be thought as the limiting case of action of gravity in stars. not only by forcing the change of state of the matter (discussed above), but also 
leading to the highest available found in nature.

The evolution of black hole in a cosmological environment has been considered before. At some point the background becomes as cool as the black hole itself, driven by the expansion of the Universe, but further evolution can not preserve the thermal equilibrium (\citealt{CustodioHorvath2003}). Rather, the black hole begins to evaporate and eventually disappears (\citealt{Hawking1975}) or leaves a tiny quantum residual, meaning that most of its entropy returns eventually to the environment.

We have seen in this work that entropy in its multifarious forms plays an important role in Stellar Evolution theory. We discussed how the gravitational contraction/collapse, although irreversible in nature {\it lowers} the entropy of matter relatively to its initial state. However, we still have a long journey towards complete understanding of the role of entropy in the fate of stars. We conjecture, however, that there is a synthesis to be made from the study of entropy in stars from a totally general point of view.

\normalem
\begin{acknowledgements}
We acknowledge the financial support received from the Funda\c
c\~ao de Amparo \`a Pesquisa do Estado de S\~ao Paulo. J.E.H.
wishes to acknowledge the CNPq Agency (Brazil) for partial
financial support.
\end{acknowledgements}

\bibliographystyle{raa}
\bibliography{bibtex}

\end{document}